\documentclass[a4paper,11pt]{article}
\pdfoutput=1 
\usepackage{jheppub,slashed,multicol}

\title{ALPs at Colliders}
\author{Ken Mimasu}
\author{and Ver\'onica Sanz}
\affiliation{
Department of Physics and Astronomy, University of Sussex, Brighton BN1 9QH, UK
}
\emailAdd{k.mimasu@sussex.ac.uk}
\emailAdd{v.sanz@sussex.ac.uk}
\date{\today}
\abstract{New pseudo-scalars, often called axion-like particles (ALPs), abound in model-building and are often associated with the breaking of a new symmetry. Traditional searches and indirect bounds are limited to light axions, typically in or below the KeV range for ALPs coupled to photons. We present collider bounds on ALPs from mono-$\gamma$, tri-$\gamma$ and mono-jet searches in a model independent fashion, as well as the prospects for the LHC and future machines. We find that they are complementary to existing searches, as they are sensitive to heavier ALPs and have the capability to cover an otherwise inaccessible region of parameter space. We also show that, assuming certain model dependent correlations between the ALP coupling to photons and gluons as well as considering the validity of the effective description of ALP interactions, mono-jet searches are in fact more suitable and effective in indirectly constraining ALP scenarios.
}

\begin{document}
\maketitle

\section{Introduction}
A long time has passed since the work of Peccei and Quinn~\cite{pq}, and yet no concrete hint of ALPs has been found. In this paper we discuss an alternative to the traditional ways of looking for ALPs using colliders. Colliders can both explore ALP masses beyond the capabilities of astrophysical constraints and in the regions they do probe, provide a cross-check exclusion limit.

In the original formulation, the Peccei-Quinn-Weinberg-Wilzcek (PQWW) axion\footnote{ Note that an alternative solution to the strong CP problem does not involve axions, but the spontaneous breaking of CP by the strong sector~\cite{annetal}.}, the scale of the interactions was related to the scale of electroweak symmetry breaking, and hence quite constrained. As this minimal realisation of a QCD axion was set aside, the idea of invisible axions, with interactions suppressed by much higher scales, arose. These invisible ALPs are not harmless, as they leave an imprint in the thermal history of the Universe. Broadly speaking, ALPs interactions are bounded from both sides: if they are too weak, ALPs could be stable and overclose the Universe, If they interact too strongly, they would affect processes such stellar formation. These cosmological/astrophysics limits depend on how well one can identify axions as the sole player in the relevant processes. For example, axions could be diluted by interacting with a hidden sector. Hence, having a variety of ways to look for ALPs provides an additional way of testing these assumptions.

In this paper, we explore a wide range of the ALP parameter space, where interactions and masses violate the symmetry responsible for the ALPs origin. A question that comes to mind, then, is whether it makes sense to rely on symmetry arguments when the symmetry is anomalous or explicitly broken. The answer depends on the ultra-violet (UV) completion of the effective ALP model. For example, the PQ symmetry could be a gauge symmetry in the UV, spontaneously broken and partly Higgssed, leading to a lighter state which obtains a potential {\it \`a la} Coleman-Weinberg. 

This paper is organized as follows. In Sec.~\ref{s:alpint} we present the Lagrangian of ALP interactions and some benchmark models. We then move onto Collider searches in Sec.~\ref{s:alpcol}, and study in a model independent fashion how to set bounds on the coupling of ALPs to photons and gluons. In the next section, Sec.~\ref{s:bound}, we obtain a combined bound on the ALP coupling to gluons and photons using the current LHC dataset, and overlay the predictions for benchmark models. Sec.~\ref{s:validity} considers the validity of the effective description of ALP interactions that we adopt in this work int eh face of the current and prospective sensitivities of collider experiments. We finalize with proposing new searches at colliders to maximize sensitivity to ALPs in Sec.~\ref{s:future} before concluding in Sec.~\ref{s:conclusions}.

\section{The origin of ALPs}\label{s:alpint}
We consider an effective Lagrangian of ALP interactions up to dimension 5 given by~\cite{Randall,kimrev}, 
\begin{align}\label{eqn:L_eff}
    \mathcal{L}_a = &\frac{1}{2}\partial_{\mu}a\,\partial^{\mu}a  - \frac{1}{2}M_{a}^2 a^2
                  -\frac{g_{a\gamma}}{4}a\,F_{\mu\nu}\tilde{F}^{\mu\nu} \nonumber \\
                   -& \frac{g_{ag}}{2}a\,\text{Tr}\left[G_{\mu\nu}\tilde{G}^{\mu\nu}\right] 
                  +\sum_{\psi}g^{\psi}_{a} \,m_\psi\,a\bar \psi \gamma^5 \psi\,,
\end{align}
where $F_{\mu\nu}$ and $G_{\mu\nu}$ are the electromagnetic and QCD field strength tensors and $m_\psi$ is the mass of the fermion $\psi$. The dimensionful couplings, $g_{a\gamma}$,  $g_{ag}$ and $g^{\psi}_{a}$, control the strength of the ALP's interactions with the gauge bosons and fermions. Other terms involving interactions with the Higgs and electroweak gauge boson field strengths are also allowed but we do not consider them here as we focus solely on searches involving photon or jet final states sensitive to this effective Lagrangian consistent with Lorentz, CP and $U(1)_{EM}\times SU(3)_c$ symmetries.

In the most general, model-independent case the couplings to photons, gluons and fermions can be considered independent and will be investigated separately 
in the context of collider physics. In specific models, though, the origin of these terms is linked. The initial model for the axion, the so-called PQWW axion~\cite{pq,pqww}, is linked to a solution to the QCD $\theta$-problem, and is very constrained by data~\cite{qcdaxout}. 
In the category of invisible axions, we find two popular axion models, DSFZ~\cite{dsfz} and  KSVZ~\cite{ksvz, newringwald}. Besides these archetypical examples, axion-like particles abound in the model-building bestiarum, e.g. axions as Dark Matter mediators~\cite{axDM}, axion Dark Matter~\cite{ADM}, axions from the compactification of extra-dimensions~\cite{XDIMax} or more general situations with multi-axion sectors~\cite{newringwald}.

Most searches have been focused on lighter ALPs based on the expectation of models like PQWW with axions in the sub-GeV range. But this is by no means a general prediction for a pseudo-Goldstone boson (PGB), as we discuss in this section. Obviously, any massive scalar particle with quantum numbers $J^{CP}=0^{-}$ would couple in a CP-conserving way as in our original Lagrangian, Eq.~\ref{eqn:L_eff}. So the question is what kind of models would lead to a CP-odd massive particle. Note that general PGBs could be either CP-even or CP-odd. This assignment of quantum numbers would depend on how they couple to fermions and how the field is embedded in the SM. For example, in Technicolor, the techni-pions~\cite{technipions} would be CP-odd and perfect candidates for ALPs, but in the Minimal Composite Higgs model~\cite{MCHM}, the resulting PGBs are CP-even (with the would-be Goldstones CP-odd) and a Higgs candidate.

A specific example is with the ALP as a member of a composite Higgs sector~\cite{Ben,Csaki-review}. In these models, $
M_a^2 \sim  \frac{y_f}{16 \pi^2}\, \left(\frac{\Lambda}{f}\right) \, \Lambda^2, 
$ 
where $y_f$ is the Yukawa coupling of a fermion $f$ and $\Lambda$ is the the scale of heavier states in the theory. With $f$ and $\Lambda$ around the electroweak-TeV scale, the mass of the ALP can be anywhere in the sub-GeV to the multi-TeV region.  Note that this ALP from Composite Higgs Models would couple to photons and gluons, but also to $W$ and $Z$ via the $SU(2)_{L}$ and hypercharge field strengths. 

In generic models, there would be no specific relation between the CP-odd particle and the scale which suppresses the dimension-five coupling, except the requirement that the Lagrangian in Eq.~\ref{eqn:L_eff} be a good effective description. In this paper we discuss limits on the effective photon and gluon couplings of a generic pseudoscalar coming from a variety of experiments with different characteristic energies. The validity of the constraints partly depend on the particular model interpretation in terms of the origin of the effective vertex. We discuss this in more detail in Sec.~\ref{s:validity}.

 \section{ALPs at colliders}~\label{s:alpcol}
\begin{figure*}
\centering
\includegraphics[width=0.3\linewidth]{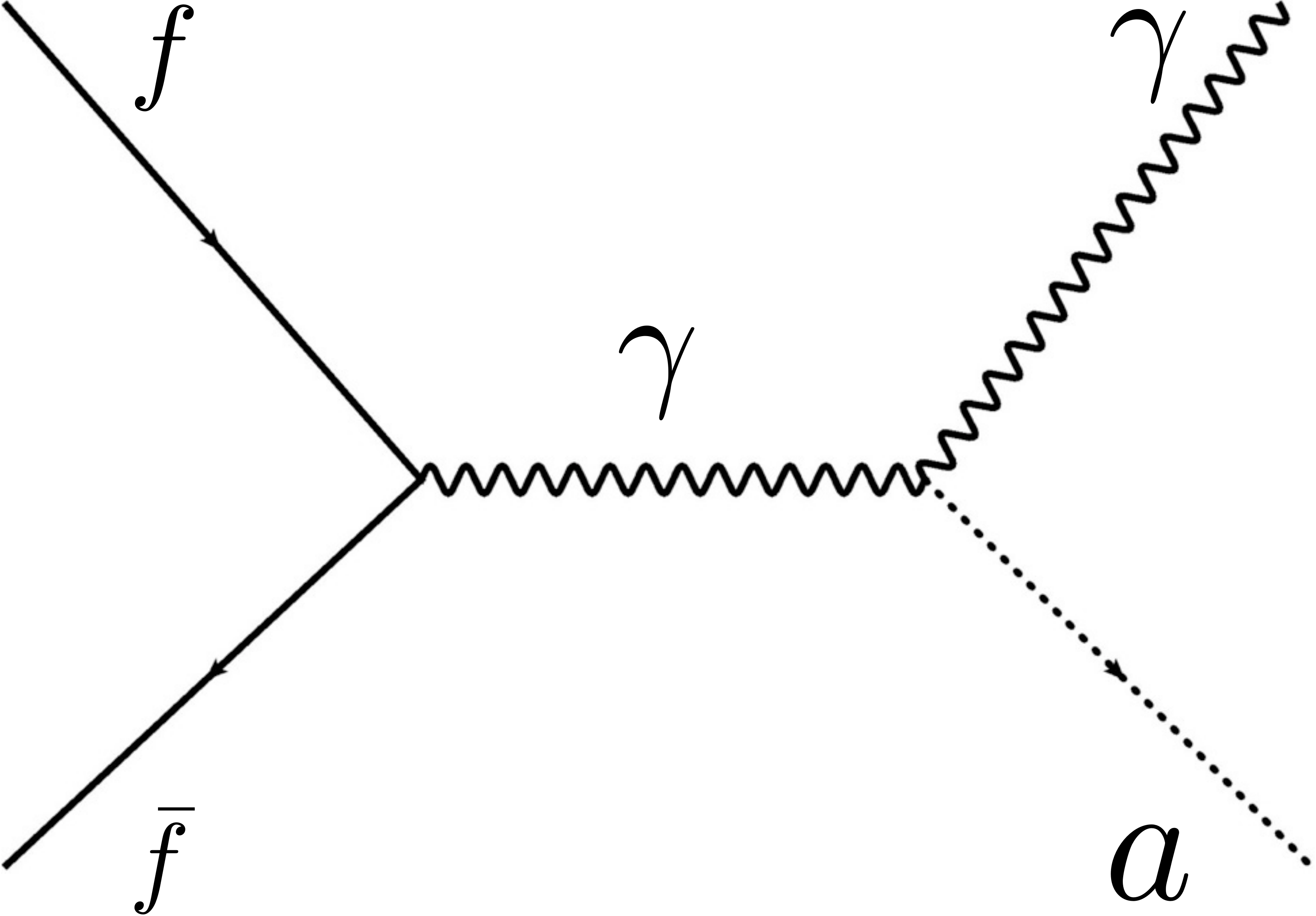}
\caption{\label{fig:ALP_prod} Feynamn diagram for ALP production in association with a photon giving rise to the mono-$\gamma$+$\slashed{E}_T$ signature.}
\end{figure*}

ALPs could be produced at colliders in association with either a photon or a jet and observed through mono-$\gamma$/jet + missing energy ($\slashed{E}_T$) channels (see~\cite{Kleban:2005rj,Jaeckel:2012yz} for related works, including limits on the couplings to electroweak gauge bosons not considered here) if the ALP is long-lived enough to decay outside of the detector volume\footnote{Note that, in the ALP as a dark matter mediator, the same signature would be obtained for prompt decays of the ALP to Dark Matter. This possibility has been explored in Ref.~\cite{HyunMin}.}.  Fig.~\ref{fig:ALP_prod} depicts the Feynman diagram for such a process. However, large enough values of $M_{a}$ or $g_{aX}$ would allow the ALP to decay within the detector and yield a tri-$X$ signature, potentially involving a displaced vertex. The partial width of an ALP to a pair of vector bosons is given by:
\begin{align}\label{eqn:ALP_width}
\Gamma_{a\to XX} = \frac{g_{aX}^2M_a^3}{64\pi}\cdot C_X,
\end{align}
where $C_X$ is a factor accounting for $X$'s colour degrees of freedom and is 1 for photons and 8 for gluons. Whether the ALP decays outside of the detector volume -- manifesting itself as missing energy,
 -- or within the volume -- producing a prompt or displaced di-$X$ pair -- depends not only on this value but also on its characteristic momentum. This is an experiment-dependent quantity which will vary with the centre of mass energy of the collider, the kinematical cuts employed in the selection and also the radius of the calorimetry system. In the following we discuss both the stable and unstable possibilities, as well as di-jet signatures where the ALP is produced via $s$-channel exchange. 

We aim to summarise current and prospective collider limits on the $(M_{a}$, $g_{aX})$ parameter space against the well known astrophysical and cosmological bounds to identify regions in which collider physics can be a complementary avenue to search for ALPs. We generated the signal using {\sc MadGraph5$\_$aMC$@$NLO}~\cite{madgraph} from a model implemented in {\sc FeynRules}~\cite{Christensen:2008py,Degrande:2011ua}. Where relevant, events were showered and hadronised with {\sc Pythia} 6~\cite{pythia} and the detector response was provided by {\sc Delphes} 3~\cite{delphes}, which makes use of {\sc FastJet}~\cite{fastjet}. Default CMS and ATLAS cards were used for {\sc Delphes}, to which the only modifications made were to the isolation and jet finding algorithm cone sizes to match those used in the analyses we reinterpret. Unless stated otherwise, the {\sc CTEQ6L1} PDFs were used.
 \subsection{Testing the coupling to photons\label{s:ALP_photon_coupling}}
\begin{figure*}
\centering
\includegraphics[width=\linewidth]{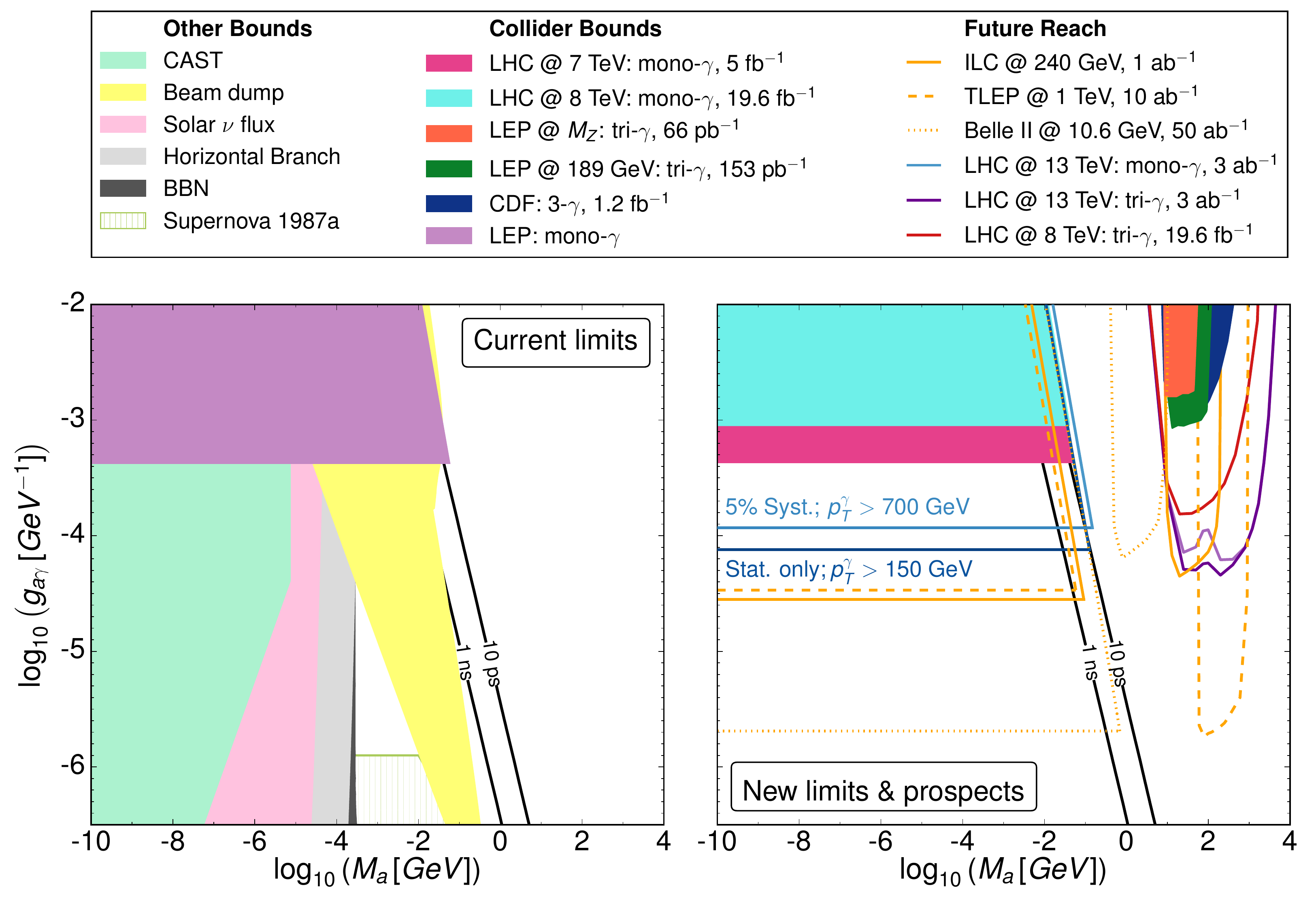}
\caption{\label{fig:ALP_gamma_limits} Current and prospective limits on ALPs in the ($M_{a}$, $g_{a\gamma}$) plane. Shaded areas represent existing experimental limits while lines denote the projected sensitivity of furture experiments.}
\end{figure*}

The ALP coupling to photons is the primary parameter through which cosmological and astrophysical bounds are set on these objects~\cite{Raffelt:2006cw}. In particular, Helioscope experiments such as CAST constrain this coupling up to a certain ALP mass by trying to convert ALPs that may have been produced by the sun into photons via the Primakoff process. Measures of the flux of neutrinos coming from the sun combined with knowledge of its temperature can also place an upper limit on additional energy loss via ALPS and thus place a bound on the coupling. Additionally, surveys of so-called `horizontal branch' stars in globular clusters also limits the coupling by its correlation with their typical helium burning lifetime. Finally, the characterisation of neutrino fluxes from supernova (SN) 1987a also provides limits~\cite{Masso:1995tw}. In all cases, the characteristic temperatures of the astrophysical processes along with the suppression of ALP-$\gamma$ conversion for large ALP masses due to energy-momentum mismatch, kinematically limit their reach as well as our ability to induce and observe a conversion. 

Aside from limits involving stellar ALP production, the presence of an ALP in the early universe would also have an effect on the Cosmic Microwave Background (CMB) and Big-Bang Nucleosynthesis (BBN) (see Ref.~\cite{Masso:1995tw,redondocadamuro} for excellent reviews). Finally, beam dump experiments are a terrestrial probe with a good sensitivity to ALP-$\gamma$ couplings. The large photon flux arising from the dump leads to a non-zero probability of conversion into and subsequent decay of ALPs for values of the masses and couplings most similar to our collider sensitivity.  

These main constraints, taken from~\cite{Hewett:2012ns}, are replicated in the left plot of Fig.~\ref{fig:ALP_gamma_limits} alongside our collider limits in the right plot. Of all of the bounds, the beam dump experiments exclude a region closest to the collider sensitivities. However, we point out the presence of a clear gap in parameter space between the sensitivities of stellar and BBN limits and that of the beam dump experiment. The fact that the beam dump experiments are sensitive to a particular range of decay lengths for these objects explains the orientation of the exclusion along lines of constant ALP lifetime, also shown in Figure~\ref{fig:ALP_gamma_limits}. Conversely, `extra-terrestrial' considerations tend to be limited largely by kinematics leading to a cut-off in the ALP mass sensitivity independent of the couplings. This opens a triangular region in the left plot, spanning roughly two orders of magnitude in mass and couplings, that has not been investigated thus far, in addition to the region to the right of the beam dump experiment limits. We will show that future collider experiments could be sensitive to these unexplored regions.
\vspace{0.2cm}
\subsubsection{Monophoton signatures\label{s:monophoton}}
Collider experiments provide a continuous bridge between the eV to KeV reaches of the astrophysical limits and higher MeV to GeV masses. The lighter masses are conducive to a stable ALP and are best searched for in mono-$\gamma$ + $\slashed{E}_T$ searches. These have been performed at the 7 and 8 TeV runs of the LHC~\cite{Chatrchyan:2012tea,Aad:2012fw,CMS:2014mea,Aad:2014tda} as well as previously at LEP~\cite{LEP_monophoton}. The limit from electron-positron colliders, having already been determined, was taken from~\cite{Hewett:2012ns} and reproduced in Fig~\ref{fig:ALP_gamma_limits} (solid purple area) while we reinterpret the LHC analyses in order to constrain the ALP scenario.

In both cases, the analyses required the presence of one high $p_T$, isolated photon and allowed for the presence of one additional soft jet, providing it was well separated from the photon and/or the $\slashed{E}_T$. Further details of the selection process in both analyses can be found in Appendix~\ref{AppendixA}. At hadron colliders, the $p_T$ cut is not only useful for suppressing backgrounds but is also necessary for predicting the signal contribution as it provides a hard scale or cut-off to reduce the sensitivity to the PDF uncertainties in the low-$x$ and/or low $Q^2$ domains. Moreover, it is also required by the presence of an $s$-channel photon in the partonic process.

We generated a sample of $p p \to a \gamma$ and $p p \to a \gamma + 1{\text{ jet}}$ events at 7 and 8 TeV, matched at a $k_T$ distance measure of 25 GeV. The cross section resulting from parton shower and matching procedures for the signal multiplied by the selection efficiency of each analysis was thus computed as a function of $M_a$ and $g_{a\gamma}$. In fact, the range of couplings to which current collider analyses are sensitive ensure that the ALP only remains collider-stable up to sub GeV masses. This means that relevant production cross sections for missing energy searches can safely assume the ALP mass to be negligible with respect to the typical energy scale of the process, set by the photon $p_T$ or the $\slashed{E}_T$ cuts. Moreover the leading order matrix element $q\bar{q}\to a \gamma$ does not depend on the centre of mass energy, $\hat{s}$, up to kinematical effects proportional to $s-M_a^2$. This removes the mass dependence and leads to a direct constraint on $g_{a\gamma}$ that scales with the integrated luminosity. 

For the 7 TeV analyses, the observation of 73(116) candidate events compared to a background expectation of 75.1$\pm$9.5(137$\pm18^{stat.}\pm9^{syst.}$) for CMS(ATLAS), dominantly arising from $Z\to\nu\nu+\gamma$, leads to a combined limit on $g_{a\gamma}$ of $4.3\times10^{-4}$ GeV$^{-1}$. In the CMS 8 TeV analysis~\cite{CMS:2014mea}, model independent limits on the cross section times acceptance, $\sigma\times\mathcal{A}$, were provided as a function of the cut on $p_T^{\gamma}$ with stringent limit set by the region with $p_T^{\gamma}>250$ GeV. This turned out to be weaker than their 7 TeV counterpart suggesting that a more tailored analysis may help to improve these bounds. The ATLAS analysis~\cite{Aad:2014tda} only has one signal region with a $p_T^\gamma$ cut of 125 GeV ($\slashed{E}_T >$ 150 GeV), and was found to set a limit 2.5 times weaker than the corresponding CMS one. We therefore do not show it in our figures.

The dark-pink and cyan shaded areas of Fig.~\ref{fig:ALP_gamma_limits} reflect the 7 and 8 TeV limits in the $(M_{a}, g_{a\gamma})$ plane respectively. The collider stability condition is imposed by demanding that the decay length of the ALP exceed the radius of the ATLAS calorimetry system~\cite{ATLAS:1999uwa} assuming it decays only into photons. A mild mass dependence enters in the relationship between the laboratory frame lifetime and the proper lifetime, via the characteristic energy of the experiment, set by the $p_T^{\gamma}$ cut. We see that the current LHC sensitivity is roughly equivalent to the previously set LEP bounds.

The prospective reach of the high energy LHC run was also considered. Assuming the full dataset of 3 ab$^{-1}$ of 13 TeV data, we simulated the main irreducible background to the search, $p p \to Z + \gamma;\,Z \to \nu\bar{\nu}$, to estimate the values of mass and coupling that could be excluded at 95\% CL. Signal to background ratios were optimised by simply cutting on the photon $p_T$ assuming an angular acceptance of $|\eta| < 2.5$. The limits are represented by the blue lines in Fig.~\ref{fig:ALP_gamma_limits}. In one case, purely statistical uncertainties were considered while in the other, a 5\% systematic uncertainty is ascribed to the background expectation. The $p_T$ cuts that gave the best exclusion were 150 and 700 GeV respectively, with the exclusion set at around $10^{-4}$ GeV$^{-1}$. Ultimately, as this is a projection for the full 13 TeV LHC dataset, the level of control over systematic uncertainties is unknown. It is therefore instructive to see the effect of assuming some level of uncertainty. As the 8 TeV analyses find a total systematic uncertainty on the expected background yields of order 10\%, it seems reasonable to assume this may be halved for an analysis of the full 3 ab$^{-1}$ of LHC run II data. The larger the background present in an analysis, the more strongly a relative uncertainty on this background affects the sensitivity. This explains the necessity of varying the $p_T$ cut between the two versions of the analysis, as cutting harder on the photon $p_T$ reduced the SM background and therefore the effect of a relative uncertainty on it.

A similar exercise can be performed for future $e^+e^-$ colliders. By virtue of the independence of the partonic matrix element to the centre of mass energy, up to kinematic effects, the sensitivity of these experiments depends only on the integrated luminosity collected. This motivated considering low energy collider experiments like Belle II (10.6 GeV), which has a very large planned integrated luminosity of 50 ab$^{-1}$, in addition to the ILC (240 GeV) and TLEP (1 TeV) experiments, which plan to collect 1 and 10 ab$^{-1}$ respectively~\cite{Abe:2010gxa,Gomez-Ceballos:2013zzn,Baer:2013cma}. In this case, a pair of cuts on the photon energy and $p_T$ were made in order to mitigate the $Z\to\nu\bar{\nu}$ background. Having a handle on the exact photon energy allows for a much more efficient suppression of the background since, in the process of interest, the ALP carries half of the collider energy. The projected sensitivites are summarised in Tab.~\ref{tab:epem_cuts}.
\begin{table}
    \centering
\begin{tabular}{|c|ccc|ccc|}
    \hline
    &\multicolumn{3}{c|}{Statistics only}&\multicolumn{3}{c|}{5\% systematics}\tabularnewline
    \hline
    Collider& $p_T$ [GeV]& $E$ [GeV]&$\log_{10}(g^{95}_{a\gamma})$& $p_T$ [GeV]& $E$ [GeV]& $\log_{10}(g^{95}_{a\gamma} )$\tabularnewline
    \hline
    ILC & 80 & 115 & -4.6& 110 & 115 & -4.5  \tabularnewline
    TLEP & 330 & 495 & -4.2 & 350 & 495 & -3.9 \tabularnewline
    Belle II & 3 & 5 & -5.7& 4 & 5 & -5.1 \tabularnewline
    \hline
    \end{tabular}
    \caption{Table summarising the energy and $p_T$ cuts performed on the dominant SM background at LO with the associated 95\% CL limit on $g_{a\gamma}$ for three future $e^+e^-$ colliders. The limit is calculated assuming both statisical uncertainties only and adding a 5\% systematic uncertainty on the background expectation. \label{tab:epem_cuts}}
\end{table}

Overall, existing collider data constrains $g_{a\gamma}$ to values around $10^{-3}$ GeV$^{-1}$ while upcoming and future experiments may be able to reach values of order $10^{-6}$ GeV$^{-1}$. Most importantly, they cover the area of parameter space untested by astrophysical/beam dump experiments with the Belle II sensitivity notably approaching the supernova 1987a limits. Collider bounds are insensitive to the modelling uncertainties that may exist with respect to star formation and solar structure and therefore provide a crucial cross check in investigating ALPs.
\subsubsection{Triphoton signatures\label{s:triphoton}}
Moving away from the mono-$\gamma$ + $\slashed{E}_T$ signature, the case where the ALP may promptly decay into a pair of photons, corresponding to a complementary region of $(M_a,g_{a\gamma})$ space, also warrants consideration. A number of LEP analyses exist on two and three photon final states testing both anomalous decay modes of the Z in the case of LEP 1 and QED predictions in that of LEP 2. The CDF collaboration also performs a search for the anomalous production of a pair of photons in association with a number of additional particles, of which one possibility is an extra photon~\cite{CDF_diphoton_plus_X}. This search mode is a subset of those proposed to search for technipions, which also come under our broad definition of ALPs~\cite{Lubicz:1995xi,Eichten:1997yq,Lane:2002wb,Zerwekh:2002ex}. For each of the LEP 1 and 2 runs, we chose to reinterpret the analysis that used the largest integrated luminosity.

We reinterpret the analyses by generating the triphoton signal at parton level and replicating the selection process as closely as possible. Since a simple three photon final state at an $e^+e^-$ collider should be relatively clean, we do not expect there to be a big difference in the signal acceptance if we were to include a simulated detector response and ISR effects. At the Z-pole, the analysis chosen was by the L3 experiment~\cite{Acciarri:1994gb}, while the higher energy run with the highest integrated luminosity came from DELPHI~\cite{Anashkin:1999da}. Precise details of each selection can be found in Appendix~\ref{AppendixB} but they can be generally summarised as requiring a pair of isolated, well identified photons above a certain energy threshold within an angular acceptance accompanied by a third photon with slightly relaxed criteria. 

An important factor in these analyses is the angular isolation requirement on the observed photons. If $M_a$ is small compared to $\sqrt{s}$, the diphoton system from the ALP decay will be highly boosted from carrying half of the centre of mass energy, corresponding to a velocity factor of $\beta = \frac{s-M_a^2}{s+M_a^2}$. This will result in a collimated pair of photons in the laboratory frame that will almost certainly fail the isolation conditions. This means that, in this channel, a given experiment is only sensitive to ALP masses around its centre of mass energy, where the ALP is produced close enough to rest that its decay products remain sufficiently separated. This is reflected in the narrow mass windows over which each search sets bounds in Figure~\ref{fig:ALP_gamma_limits}. The limits were found in a similar way to the monophoton case comparing the 7(7)[4] observed events to the background expectations of $7.1\pm0.7$($9.6\pm0.5$)[$2.2\pm0.6$] for the L3(DELPHI)[CDF] analyses and extracting the 95\% CL exclusions as a function of $M_a$ and $g_{a\gamma}$. 

The case of future $e^+e^-$ colliders was also investigated as with the mono-$\gamma$ searches, taking into account the main irreducible background of $e^+e^-\to 3\gamma$. This was simulated for the three colliders considered in the previous section and passed through a simple analysis exploiting the differences in kinematics between the signal and background. Since the sensitivity of the experiment relies on the ALP having a significant mass compared to $\sqrt{s}$, the energy of the recoiling photon will be $\frac{s-M_a^2}{2\sqrt{s}}$. Futhermore, a given $M_a$ also predicts a minimum angular separation between the decay photons of the ALP in the collider frame, given by $\cos\theta_\gamma = 2\beta^2-1=1-\frac{8s M_a^2}{(s+M_a^2)^2}$. Finally, the invariant mass of the decay pair can also be required to lie close to $M_a$. Combining this information into a `cut and count' procedure, implementing basic isolation and acceptance cuts summarised in Appendix~\ref{ss:future_tri} yields the projectied limits shown on the right hand side of Fig.~\ref{fig:ALP_gamma_limits}. Once again, the $p^2$ dependence of the ALP-$\gamma$ interaction and the consequent $\sqrt{s}$ independence of the production process plays a key role in determining the sensitivities of each experiment. While Belle II will supply the greatest integrated luminosity, the SM background is much larger in its energy range ($\sim$pb) and the sensitivity is therefore reduced. This will particlarly be true if systematics are taken into account. Conversely, the SM tri-$\gamma$ cross sections at TLEP are of order 0.1 fb post-selection and lead to much better limits, reaching the same order of magnitude in sensitivity to the Belle II monophoton analysis.  

Finally we also consider the LHC prospects for this signature, both for the 8 and 13 TeV runs. In this case, another potential source of backgrounds exists in the form of diphoton + 1 jet where the jet fakes a photon. We therefore simulated both this and the irreducible SM triphoton background and followed a similar approach to the linear collider case, analysing the signal and background events as a function of $M_a$. The details of the analysis can be found in Appendix~\ref{ss:future_tri} but can be summarised by the identification of two candidate decay photons, whose invariant mass most closely reconstructs $M_a$, and a remaining `recoiling' photon. After basic acceptance requirements, the invariant mass of the decay pair was required to lie within 10\% of $M_a$, down to a minimum resolution of 5 GeV for the lower mass cases. The energy of the recoiling photon in the three photon centre of mass frame was cut upon as a function of $M_a$. The misidentification rate for a jet faking a photon was taken to be 10$^{-3}$, which is seen as relatively conservative estaimate for jets with $p_T>$ 20 or 30 GeV (see for example~\cite{Lilley:2011mea}). In any case, the irreducible SM background was found to always be more significant than its reducible counterpart. The results of the analyses are also shown in Fig.~\ref{fig:ALP_gamma_limits}, demonstrating a similar sensitivity of the tri-photon analyses to their mono-photon equivalents. Both analyses are sensitive to ALP masses between 50 and 1000 GeV with the 8 TeV (dark red line) probing at best $g_{a\gamma}\sim 10^{-3.8}$ GeV$^{-1}$ and the 13 TeV (purple lines), $g_{a\gamma}\sim 10^{-4.4}$ GeV$^{-1}$. For reasons discussed in Sec.~\ref{s:monophoton}, we quote sensitivities assuming statistical uncertainties only as well as a 5\% systematic uncertainty on the total background expectation (dark and light purple lines, respectively) added in quadrature with the quoted renormalisation and factorisation scale uncertainties of the subdominant diphoton at NLO sample (typically of order 20\%). These were only relevant for 13 TeV as the 8 TeV analyses were statistically dominated. 

\subsubsection{Displaced vertices} 
In the region where the ALP yields a displaced vertex, it will decay into two close-by or even overlapping photons. Although there are currently no searches sensitive to this signature, the high energy run of the LHC may by able to search for such objects. For example, a combination of shower shape and timing information could be used to disentangle a calorimeter deposition resulting from a pair of collimated photons or a displaced vertex. If the ALP decay length is such that no sizeable deposition on the electromagnetic calorimeter is left, this configuration would revert to the mono-$\gamma$ plus missing energy situation described before.

\subsection{Testing the coupling to gluons\label{s:ALP_gluon_coupling}}
Existing 8 TeV LHC mono-jet analyses were used to set limits in an analogous parameter plane to the photon case, but for $g_{ag}$. The process is more involved than the monophoton one 
owing to the larger number of diagrams contributing to the process and the coloured final state. Of all of the possible diagram 
topologies involving both quark and gluon initial and final states, the dominant diagram was found to be $g g \to a\,g$. 
The presence of a jet in the final state motivated a detailed study including parton shower and hadronisation effects, as in the monophoton case. Events for the process $pp\to a j$ were generated for the 8 TeV LHC, where we observed the mass independence of the signal in the regions of couplings for which the ALP remained collider stable assuming only decays to gluons. Given the factor 8 increase in the decay width of the ALP into gluons due to the colour degrees of freedom, the collider stability limit is reached earlier compared to photon decays. 

\begin{figure*}
\centering
\includegraphics[width=0.4\linewidth]{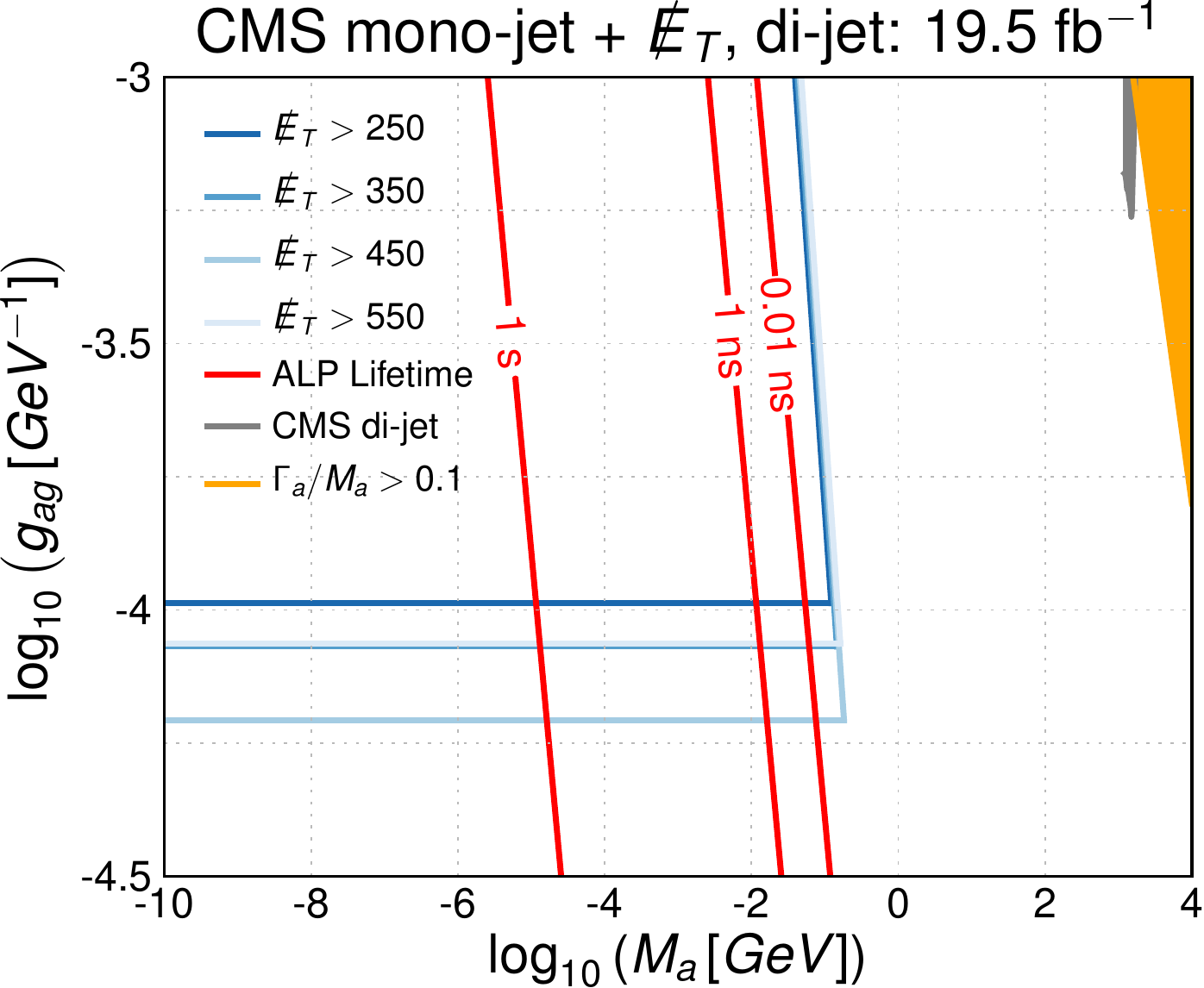}
\includegraphics[width=0.4\linewidth]{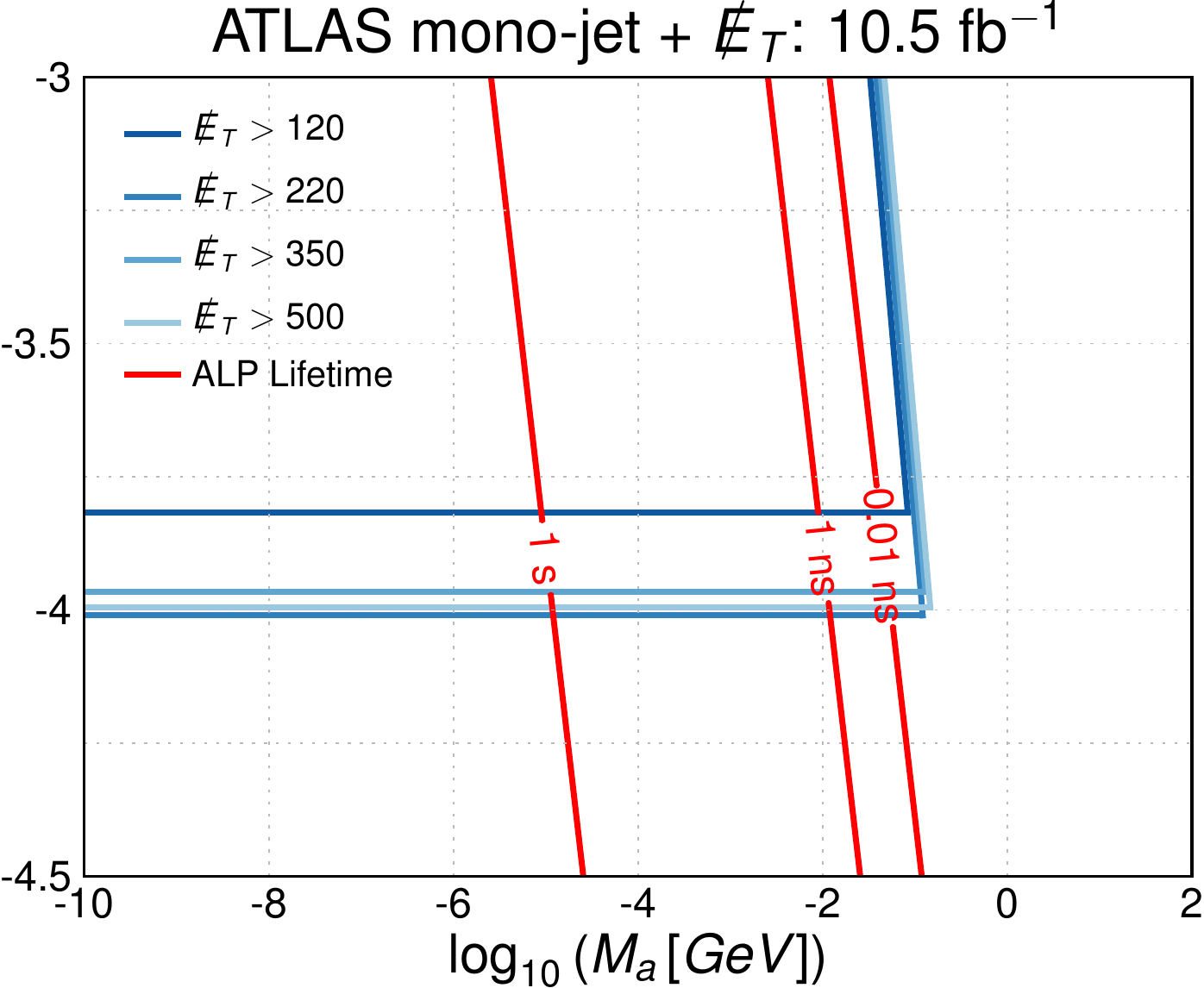}
\caption{\label{fig:CMS_monojet_limits} Limits in the ($M_{ax}$, $g^{g}_{ax}$) derived from the seven signal regions in 
         the current CMS and ATLAS mono-jet analyses. Refer to Appendix~\ref{AppendixA} for a definition of the selection processes and signal regions}
\end{figure*}

The generated events were passed through the selection procedures of the latest CMS~\cite{CMS:rwa} and ATLAS~\cite{ATLAS:2012zim} 
monojet analyses which are summarised in Appendix~~\ref{AppendixB}. In both cases a basic monojet selection is performed, requiring exactly 
one hard jet and vetoing events containing leptons or a second jet above a low $p_T$ threshold. The events are categorised into 
several signal regions defined by a $\slashed{E}_T$ cut ranging from 120 to 550 GeV. Figure~\ref{fig:CMS_monojet_limits} shows 
the excluded regions in parameter space of the CMS analysis for each signal region showing a sensitivity to couplings of order
$g_{ag}\sim 10 ^{-4}$ GeV$^{-1}$. 

 \subsection{Alternative searches}\label{s:alternative}
In the region of heavier ALPs, with the ALP decaying near the interaction vertex, one would require a dedicated search looking for a signature of three jets, with two of them reconstructing a resonance~\cite{AndiW}. An interesting alternative would also be the case where the ALP is produced on-shell via gluon-gluon fusion and subsequently decays into a pair of jets. We attempt to assess the sensitivity of such searches to high-mass ALPs by considering limits on the cross section $\times$ acceptance $\times$ branching ratio imposed by a search for dijet resonances by CMS~\cite{CMS:kxa}. This was done by computing the production cross section times the branching ratio as a function of ALP mass and coupling, multiplying by the quoted approximate acceptance factor of 0.6 and comparing with the excluded cross section provided. Limits are given in Figure~\ref{fig:CMS_monojet_limits}, showing a limited sensitivity to couplings of order $10^{-3}$ GeV$^{-1}$ in a range of masses between 1-3 TeV. As mentioned in the beginning of this section, the experimental search targets resonances narrower than the dijet mass resolution. We therefore choose to limit the capabilities of this search to ALP widths below 10\% of the mass as reflected by the orange region in the upper corner of the left figure.

 \subsection{Combined limits and model-dependent correlations}\label{s:bound}
\begin{figure}
\centering
\includegraphics[height=0.222\textheight]{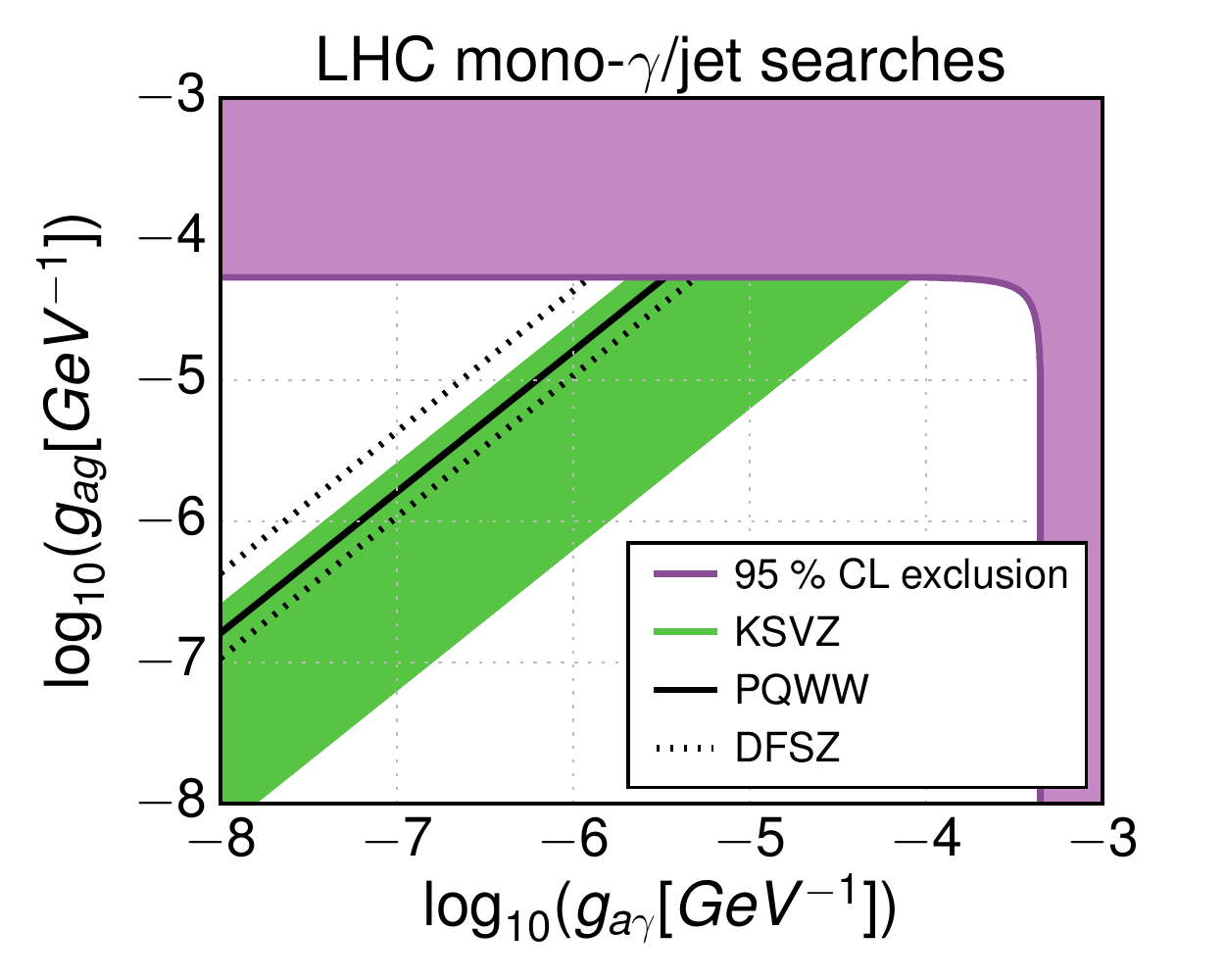}
\includegraphics[height=0.218\textheight]{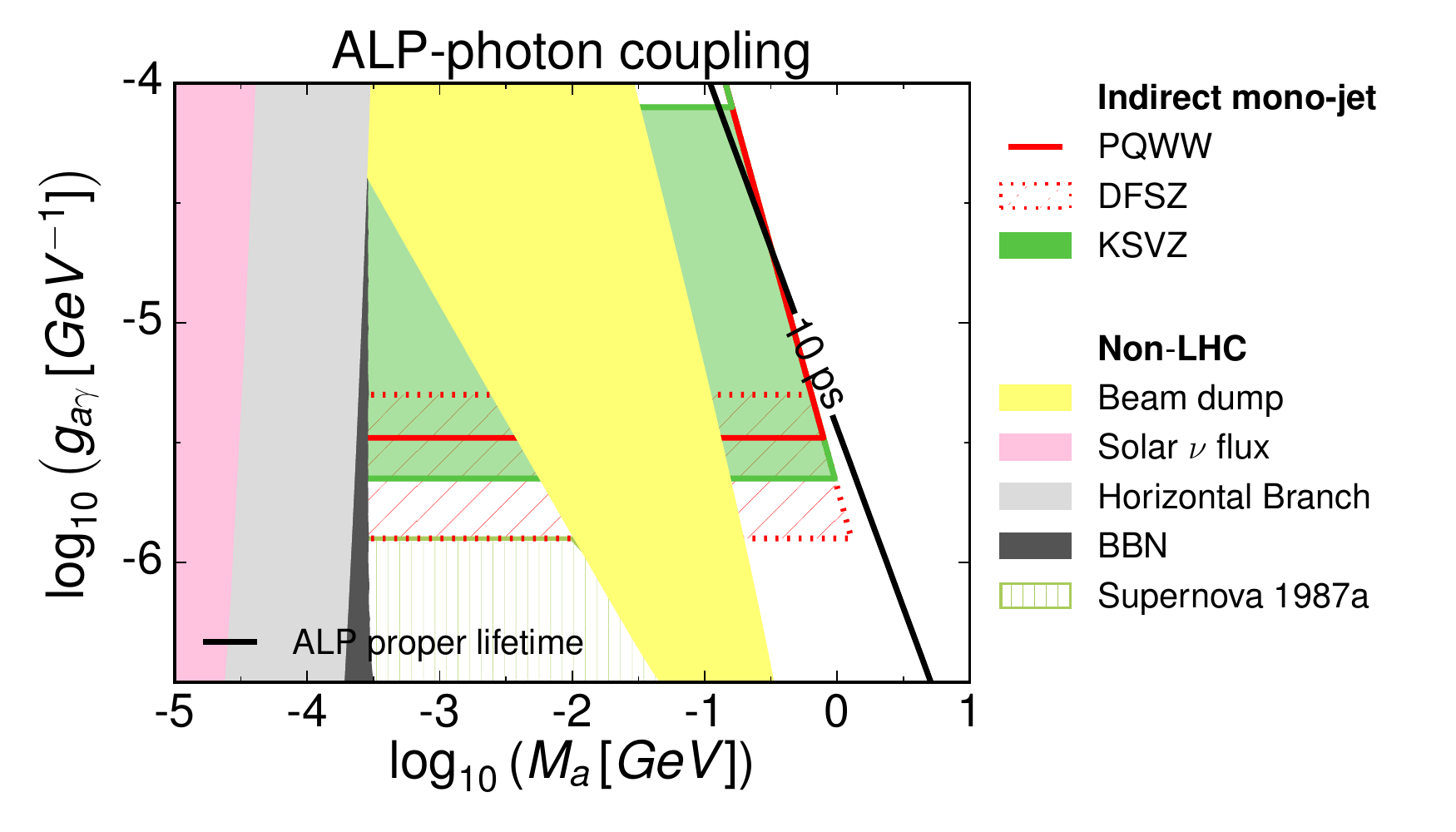}
\caption{\label{fig:ALP_combined_limits}\emph{Left}: Present limits on the coupling of ALPS to photons and gluons (purple area) along with correlations between the two couplings in different models (black lines and green band). \emph{Right}: An illustration of the indirect limits on $g_{a\gamma}$ set by the monojet search assmuing the model dependent correlations.}
\end{figure}

We combine constraints from both mono-$\gamma$ and mono-jet signatures in a $\chi^2$ fit to the ALP  parameters, by exploiting the fact that  searches are largely independent of the value of $M_a$, up to values of 10$^{-1}$--10$^{-2}$ GeV. In the mono-jet case, given that the different signal regions are all correlated, we refrained from combining them but selected the most constraining one from each analysis. The combination is shown in the left hand plot of Fig.~\ref{fig:ALP_combined_limits}.
In specific models, there are relations between the ALP coupling to photons and gluons, and these can be exploited to set stronger bounds. In the same figure, we show the correlations which appear in the QCD axion models mentioned in the first section: PQWW, KSVZ and DFSZ. The green band is obtained by varying the heavy quark charge of the KSVZ model while the two DFSZ lines correspond to a discrete choice over Higgs doublet couplings to leptons. Since the gluon coupling is constrained more tightly than that of the photon, in these models, mono-jet signatures constrain a value of $g_{a\gamma}$ two orders of magnitude lower than those to which current mono-$\gamma$ searches are sensitive, as shown in the right hand plot of Fig.~\ref{fig:ALP_combined_limits}. In particular one can see that the limits access a region of parameter space to which other experiments are blind (the triangular region mentioned in Sec~\ref{s:ALP_photon_coupling}) -- between the BBN and the beam bump limits as well as to the right of the beam dump limits. While such heavy ALP masses may not be favoured for these particular models, this does serve as an illustration of the kind of indirect constraints that one can obtain. As discussed in the next section, the validity of the effective interaction that we use is called into question in when interpreting the results of the photon searches in terms of these types of models. It is therefore very useful that the monojet limits can actually provide indirect constraints on $g_{a\gamma}$ from analyses where the effective theory is more reliable.

\section{Validity of the effective description}\label{s:validity}
\begin{figure}
\centering
\includegraphics[width=0.8\linewidth]{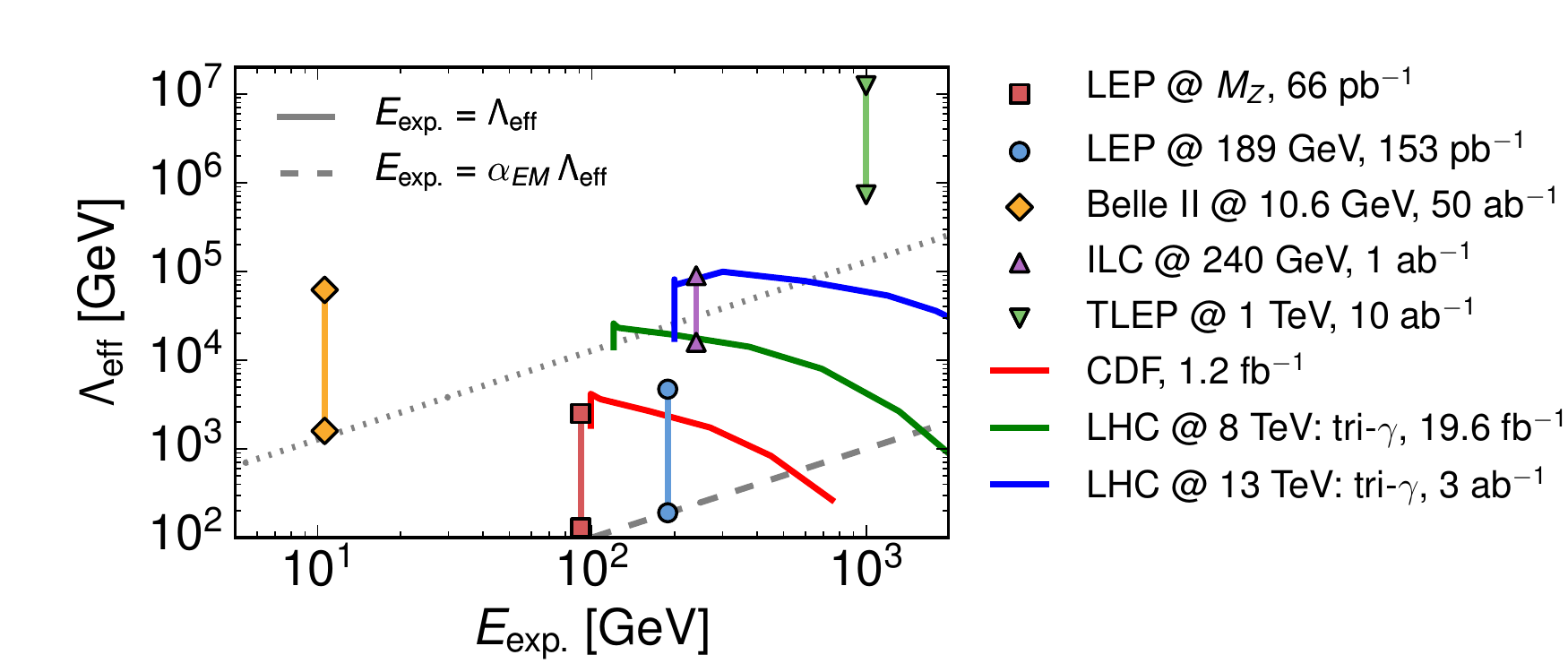}
\caption{\label{fig:validity_triphoton} Ranges of effective cutoff derived from current and prospective limits on $g_{a\gamma}$ determined in this paper from tri-$\gamma$ searches, as shown in Fig~\ref{fig:ALP_gamma_limits}. In the case of hadron colliders, the characteristic energy of each experiment is set either by $M_a$ or by the minimum $p_T$ requirement on the photons since the ALP is required to be produced nearly at rest. The scale of the $e^+e^-$ colliders is simply set by the centre of mass energy. }
\end{figure}

As mentioned in the introductory portion, the aim of this paper is to adopt an agnostic approach by parametrising generic effective interactions of ALPs to photons and gluons dictated solely by Lorentz, CP and $U(1)_{EM}\times SU(3)_c$ invariance. In this model-independent approach, these interactions appear at leading order as a non-renormalizable interaction, suppressed by a mass scale. One can express this more explicitly by defining a scale $\Lambda$ such that our couplings $g_{aX}$ scales as

\begin{align*}
   \frac{g_{aX}}{4} = \frac{c_{X}}{\Lambda}.
\end{align*}
As this theory can only be understood as an effective theory, it is important to ensure that the characteristic energies, $E_{\text{exp}}^i$, of each different experiment used to set limits on these couplings do not exceed the cutoff of the effective interaction implied by the value of $g^{i}_{aX}$ to which it is sensitive.  For $c_X\sim 1$, as long as these energies are sufficiently small compared to $\Lambda_{i}$, the predictions for the signal rates can be trusted as reasonable estimates. The energy scales of the mono-$X$ analyses are characterised by the corresponding $p_T$ or $\slashed{E}_T$ cuts that they impose. Given the fact that they generally require central photons/jets, this can be approximated as $E_{\text{exp}}\simeq 2\slashed{E}^{\text{cut}}_T$. In the case of the decaying ALPs in the higher mass region, the minimum energy is set by $M_a$ itself. Given that they are required to be produced approximately at rest in order for calorimetry to resolve their decay products, the total energy should be of this order, perhaps a few times $M_a$. We considered the differential distributions with respect to invariant mass to determine the typical energy scale of each signal region. In some cases, the basic $p_T$ requirements on the photons increased the characteristic energy beyond $M_a$ and this was taken into account in Fig.~\ref{fig:validity_triphoton}. For $e^+ e^-$ colliders, the energy scale is clearly the centre of mass energy of the machine.

Considering the values of $g_{a\gamma}$ and $g_{ag}$ on which we set/project limits in our analyses, we find that the corresponding $\Lambda$s are in the multi-TeV region, as shown in Tabs.~\ref{tab:validity_monophoton} and~\ref{tab:validity_monojet} and Fig.~\ref{fig:validity_triphoton}. As an illustration of the `worst case', the least sensitive mono-$\gamma$ experiments are those which will correspond to the lowest cutoff. This would be the 8 TeV monophoton analysis at the LHC which sets a limit of order $g_{a\gamma}\simeq 10^{-3}$. The corresponding cutoff is then $\Lambda\simeq 4.5$ TeV, which, when compared to the characteristic energy set by the $\slashed{E}_T$ cut of 250 GeV lying the the range 0.5-1.1 TeV (accounting for the maximum rapidity allowed by the $\eta$ acceptance) suggests that one is still below the cutoff but is approaching the breakdown of the EFT. All other cases respect this validity requirement more comfortably. Conversely the most sensitive future experiment, Belle II, projects a limit that corresponds to an effective theory of order PeV, compared to its relatively small centre of mass energy of 10.6 GeV. In the case of the tri-$\gamma$ searches, the lowest mass regions up to the peaks of coupling sensitivity have their energy scale set by the $p_T$ requirements of the experiment, characterised by the vertical segment of each line. In each case, the masses and energies lie well within the effective description. However, as the mass (and therefore $E_{\text{exp}}$) increases and the sensitivity goes down, the effective picture breaks down and one should not rely on such limits. The breakdown of the effective theory in this figure occurs below the dashed grey line for which the experimental energy is above the naive cutoff of the theory.

\begin{table}
    \centering
    \begin{tabular}{|c|c|c|c|}
        \hline
        Analysis&$E_{\text{exp}}$ [TeV]&$\Lambda_{\text{eff}} $ [TeV] &$\log_{10}(g^{95}_{a\gamma}\,$[GeV]$)$\tabularnewline
        \hline
        LHC7 & 0.26--2.6 & 9.4       & $-3.37$   \tabularnewline
        LHC8 & 0.5--1.1 & 4.5         & $-3.05$  \tabularnewline
        LHC14 (Stat) & 0.3--1.8 & 50& $-4.12$   \tabularnewline
        LHC14 (5\% Syst.)      & 1.4--8.6  & 34& $-3.93$  \tabularnewline
        ILC & 0.24            & 140  & $-4.47$  \tabularnewline
        TLEP & 1              & 118   & $-4.55$ \tabularnewline
        Belle II & 0.0106     & 1960  & $-5.69$ \tabularnewline
        \hline
    \end{tabular}
    \caption{\label{tab:validity_monophoton} Table summarising the effective cutoffs derived from current and prospective limits on $g_{a\gamma}$ determined in this paper from mono-$\gamma$ searches, as shown in Fig~\ref{fig:ALP_gamma_limits}. For the hadron collider analyses, the lower range of the characteristic energy of the experiment, $E_{\text{exp}}$, is determined by twice the $p_T$ or $\slashed{E}_T$ cut used. This is the dominant part of phase space in which the signal cross section lies. The higher end of $E_{\text{exp}}$ assumes the most extreme value of pseudorapidity allowed by the acceptance cut imposed i.e. $2p^{\text{cut}}_T\cosh(|\eta^{\text{cut}}|)$. }
\end{table}

\begin{table}
    \centering
    \begin{tabular}{|c|c|c|c|}
        \hline
        Analysis&$E_{\text{exp}}$ [TeV]&$\Lambda_{\text{eff}} $ [TeV] &$\log_{10}(g^{95}_{ag}\,$[GeV]$)$\tabularnewline
        \hline
        CMS&-&-&-\tabularnewline
        $\slashed{E}_T > 250 GeV  $& 0.5--2.8 & 36& $-3.95$   \tabularnewline
        $\slashed{E}_T > 350 GeV  $& 0.7--3.9 & 50& $-4.1$    \tabularnewline
        $\slashed{E}_T > 450 GeV  $& 0.9--5   & 63& $-4.2$    \tabularnewline
        $\slashed{E}_T > 550 GeV  $& 1.1--6.1 & 50& $-4.1$    \tabularnewline
        ATLAS&-&-&-\tabularnewline
        $\slashed{E}_T > 120 GeV$  & 0.24--0.9& 28 & $-3.85$   \tabularnewline
        $\slashed{E}_T > 220 GeV$  & 0.44--1.7& 40 & $-4.0$    \tabularnewline
        $\slashed{E}_T > 350 GeV$  & 0.7--2.6 & 31 & $-3.9$    \tabularnewline
        $\slashed{E}_T > 500 GeV$  & 1.0--3.8 & 40 & $-4.0$    \tabularnewline
        \hline
    \end{tabular}
    \caption{\label{tab:validity_monojet} Table summarising the effective cutoffs derived from current LHC limits on $g_{ag}$ determined in this paper from mono-jet searches, as shown in Fig.~\ref{fig:CMS_monojet_limits}. The range of characteristic energies is determined as in Tab.~\ref{tab:validity_monophoton}. The dashed and dotted lines delimit the region in which the EFT picture breaks down in that the characteristic energy of the experiment exceeds the naive cutoff of the theory. The two cases correspond to the naive cutoff, set by the dimensionful coupling $g_{a\gamma}$ and the QCD axion-like interpretation where the cutoff is reduced by a factor $\alpha_{EM}$ as discussed in the text.}
\end{table}

Tabs.~\ref{tab:validity_monophoton},~\ref{tab:validity_monojet} and Fig.~\ref{fig:validity_triphoton} show that, in the simplest effective theory interpretation of ALP interactions, the various limits shown in Figs.~\ref{fig:ALP_gamma_limits} and ~\ref{fig:CMS_monojet_limits} can be taken at face value. However, if additional assumptions about the origin of the coupling are made, one should interpret the bounds with care. If, for example, one assumes that the ALP is a PGB and that its $F\tilde{F}$ interaction originates from a trace anomaly, one would typically expect it to have a small Wilson coefficient of order a loop factor times a gauge coupling squared, 
\begin{align*}
    \frac{g_{aX}}{4}\sim\frac{\alpha_X}{4\pi f_a},
\end{align*}
where $f_a$ is the symmetry breaking scale and $\alpha_{X}$ is the fine structure constant associated to the gauge boson, $X$. In this description, the linear Lagrangian we study would be insufficient at energies around the cutoff, $\Lambda \simeq 4 \pi f_a$~\cite{georgi}, meaning that the effective cutoff of such theories is reduced by a factor $\alpha$ with respect to those previously discussed. This particularly poses problems for interpreting the mono and tri-$\gamma$ searches, given the small value of the fine structure constant. The current constraints shown in Fig.~\ref{fig:validity_triphoton}, for example, partly lie outside of the region of validity of this effective description, delimited by the dotted grey line. While the CDF limits completely lie bewloe th vail region, the LHC limits remain valid for masses between 20-100 GeV for the 8 TeV projection and 20-500 GeV for the 13 TeV case. 

As shown in Sec.~\ref{s:bound}, in this picture, models tend to have a hierarchy between $g_{ag}$ and $g_{a\gamma}$ via the ratio $\alpha_{EM}/\alpha_{S}$ which means that limits set on $g_{ag}$ from mono-jet searches can be interpreted as model-dependent, indirect limits on $g_{a\gamma}$. Tab.~\ref{tab:validity_monojet} indicates that the most sensitive signal region that dominantly contributes to the $g_{ag}$ limit plotted in Fig.~\ref{fig:ALP_combined_limits} (CMS, $\slashed{E}_T <450$ GeV) has an associated effective cutoff of 63 TeV. Anticipating that the UV completion of this effective vertex arises from a loop induced anomaly term reduces this cutoff by $\alpha_S$ to $\Lambda\sim6.3$ TeV, with the characteristic energy of this signal region lying just about within this range. This is in opposition to the complementary mono-$\gamma$ searches, where the larger reduction of the cutoff bring most current limits outside of the valid region. In this sense, having a particular model in mind, it can be argued that mono-jet searches probing $g_{ag}$ are more appropriate ways to constrain such models at present, given the current and prospective sensitivity at the LHC to $g_{a\gamma}$. It does not appear that $g_{a\gamma}$ can be constrained directly in the context of such models using existing analyses although the LHC has the potential to probe certain mass windows in the tri-photon topology. The next generation of linear colliders will be able to constrain this coupling both in the sub GeV region through to masses of order 1 TeV.
 \section{Improving searches for ALPs at colliders}\label{s:future}
So far we have based our limits and future prospects on existing searches, namely mono-X, di-jet and tri-$\gamma$ signatures. It is possible that existing data may be more sensitive to the presence of ALPs coupling to either photons and/or gluons if dedicated searches were performed. This is particularly true for the non-collider stable region of parameter space where the ALP decays within the calorimetry yielding a tri-$X$ signature with two photons or jets potentially coming from a displaced vertex. The background predictions for three photons have been calculated in various works at NLO, see, for example Refs.~\cite{Campbell:2014yka,Mandal:2014vpa}. Ultimately, one seeks large integrated luminosities to set limits on ALP interactions which makes the case for high lumisosity colliders and, as we have seen, strengthens the case for lower energy experiments such a Belle II.

Also the production of an ALP with any appreciable boost will lead to collimated decay products which could be interpreted as a single photon or jet. ALP signatures of this kind would mimic di-$X$ searches and and constraints might be obtained by reinterpreting Higgs to di-$\gamma$ searches, see e.g. Ref~\cite{ATLAS:2012soa}. A similar analysis targeting ALP production could be imagined as well as those focusing one or both of the displaced vertex and boosted decay products.

Finally considering the discussion in Sec.~\ref{s:validity}, it seems that, through direct searches with photon final states, the LHC may be sensitive to values of $g_{a\gamma}$ of order $10^{-3}$--$10^{-4}$ GeV$^{-1}$. Depending on the model interpretation, this may point to naive cutoffs below the characteristic energy of the experiment, signalling a breakdown of our effective description. In particular, QCD axion-like interpretations where the ALP-$\gamma$ coupling is induced by an axial anomaly will reduce the effective cutoff by a factor of $\alpha_{EM}$ while for mono-jet searches, the punishment of $\alpha_{S}$ is not as severe. Therefore, focussing on searches involving jets, which probe the ALP-gluon coupling may be more suitable to constrian such models indirectly by exploiting the relationship between it and the ALP-$\gamma$ coupling.
 \section{Conclusions}\label{s:conclusions}
 
Traditional searches for Axion-like particles (ALPs) span orders of magnitude in ALP interaction strength but have to focus on the low mass region. In this paper we have shown how present and future colliders are able to cover the low mass region and extend the sensitivity to larger masses until the TeV range.
 
We have studied limits on ALPs in a model-independent fashion, by switching on one type of coupling at a time. We presented current collider bounds using an array of searches involving photons, jets and missing energy. We also estimated the sensitivity of future colliders to ALPs. We found that there is a complementarity between colliders and other searches but, more importantly, future colliders will be able to close a region of the parameter space which would be inaccessible to, for example, helioscopes and existing beam dump experiments. 

We also performed a model dependent combination of signatures involving the coupling to gluons and photons, illustrating the gain of studies within specific models. The gluon coupling was found to be a safer way to indirectly constrain the photon coupling when considering the validity for th effective description. We also suggested ways for colliders to increase their sensitivity using a combination of techniques, including displaced vertices and boosted photon pairs. 
 
 \section*{Acknowledgements}
 We would like to thank Guenther Dissertori for his guidance while working on this project, as well as S. Rahatlou and D. del Re for very helpful conversations on displaced signatures. We are also very thankful to Javier Redondo for his help on clarifying aspects of non-collider searches and bounds for ALPs. This work is supported by the Science and Technology Facilities Council (grant number  ST/J000477/1). 
 
\appendix
\section{CMS and ATLAS monophoton selections\label{AppendixA}}
A summary of the main criteria of two LHC monophoton analyses used to set limits on the ALP mass and its coupling to photons using the 7 TeV data and one using the 8 TeV data. The signal samples for $p p \to a \gamma$ and $p p \to a \gamma + 1\,\text{jet}$ were matched using the MLM method in {\sc MadGraph5\_aMC@NLO} at a scale of 25 GeV. The reader should refer to~\cite{Chatrchyan:2012tea,Aad:2012fw,CMS:2014mea,Aad:2014tda} for more details on the individual analyses.
\subsection{CMS: $\sqrt{s} = 7$ TeV, 5 fb$^{-1}$}
\subsubsection*{Vetos}
    \begin{itemize}
    \item One additional jet if $p_T > 40$ GeV, $|\eta| < 3$ and distance to photon $\Delta R(\gamma,j)>0.5$ 
    \item Any number of further jets
    \end{itemize}
\subsubsection*{Selection}
    \begin{itemize}
    \item One isolated photon within a cone of $\Delta R$=0.4, $H/E <0.05$ within $|\eta| < 3$
    \item $\slashed{E}_T > 130$ GeV
    \end{itemize}

\subsection{ATLAS: $\sqrt{s} = 7$ TeV, 4.6 fb$^{-1}$}
\subsubsection*{Vetos}
    \begin{itemize}
    \item More than one jet with $p_T > 30$ GeV, $|\eta| < 4.5$
    \item Any electrons(muons) of $p_T > 20(10)$ GeV, $|\eta| < 2.47(2.4)$
    \end{itemize}
\subsubsection*{Selection}
    \begin{itemize}
    \item One isolated photon within a cone of $\Delta R$=0.4 within $|\eta| < 2.37$, excluding $|\eta|\subset[1.37,1.52]$   
    \item Photon $p_T > 150$ GeV and $\slashed{E}_T > 150$ GeV 
    \item Additional $\slashed{E}_T$ and leading jet $p_T$ cut for each signal region
    \item $\Delta\phi(\gamma,\slashed{E}_T) > 0.4$
    \item Non-vetoed jets should satisfy $\Delta R(j,\gamma) > 0.4$ and $\Delta\phi(j,\slashed{E}_T) > 0.4$
    \end{itemize}

\subsection{CMS: $\sqrt{s} = 8$ TeV, 19.6 fb$^{-1}$}
\subsubsection*{Vetos}
    \begin{itemize}
    \item More than one additional jet if $p_T > 30$ GeV with $\Delta R(\gamma,j)>0.5$ 
    \end{itemize}
\subsubsection*{Selection}
    \begin{itemize}
    \item One isolated photon within a cone of $\Delta R$=0.3, $H/E <0.05$ within $|\eta| < 1.442$
    \item $\slashed{E}_T > 140$ GeV, $\Delta\phi(\slashed{E}_T,\gamma) > 2.$
    \item $p_T >$ \{140,160,190,250,400,700\} GeV
    \end{itemize}

\subsection{ATLAS: $\sqrt{s} = 8$ TeV, 20.3 fb$^{-1}$}
\subsubsection*{Vetos}
    \begin{itemize}
    \item More than one jet
    \item Electron (muon) with $p_T>$ 7(6) GeV and $|\eta| < $ 2.47(2.5)
    \end{itemize}
\subsubsection*{Selection}
    \begin{itemize}
    \item One isolated photon (cone of $\Delta R$=0.4), $p_T >$ 125 GeV within $|\eta| < 1.37$
    \item $\slashed{E}_T > 150$ GeV, $\Delta\phi(\slashed{E}_T,\gamma) > 0.4.$
    \item Allow one additional jet if $\Delta R(\gamma,j)>0.2$ and $\Delta \phi(\gamma,j)>0.4$
    \end{itemize}

\section{L3, DELPHI and CDF triphoton selections\label{AppendixB}}
A summary of the main criteria of the LEP and CDF triphoton analyses used to set limits on the ALP mass and its coupling to photons. See Refs.~\cite{Acciarri:1994gb,Anashkin:1999da,CDF_diphoton_plus_X} for more details. The fact that our analysis was performed on a signal generated at parton level, without detector simulation means that some of the requirements are automatically satisfied. We only include the selection cuts that would affect out signal sample, designed to match the detector level requirements as closely as possible within the constraints of the information available at parton level. For example, isolation requirements  within a certain angular cone were translated to angular separation requirements given the fact that there were only ever three photons in the final state.

\subsection{L3: $\sqrt{s} = M_Z$, 65.8 pb$^{-1}$}
    \begin{itemize}
    \item Three photons within an angular acceptance of $16.1^{\circ}~<~\theta_{\gamma} < 163.9^{\circ}$ and energy $E_{\gamma}> 2$ GeV
    \item All photons to have an angular separation larger than $20^{\circ}$
    \item Additional angular restriction of $|\cos\theta_{\gamma}<0.75|$
    \item energy of the softest photon $E_{\gamma 3}/\sqrt{s}>0.125$
    \end{itemize}

\subsection{DELPHI: $\sqrt{s} = 189$ GeV, 153 pb$^{-1}$}
    \begin{itemize}
    \item Angular acceptance $\theta_{\gamma} \subset[25^{\circ},35^{\circ}]$ or $[42^{\circ},88^{\circ}]$ for the forward region and $\subset[93^{\circ},138^{\circ}]$ or $[145^{\circ},155^{\circ}]$ for the central region
    \item Two photons with $E_{\gamma}/\sqrt{s} > 0.15$ and angular separation greater than $30^{\circ}$
    \item A third photon with $E_{\gamma}/\sqrt{s} > 0.06$ isolated from the other by at least $15^{\circ}$ 
    \end{itemize}

\subsection{CDF: $\sqrt{s} = 1.8$ TeV, 1.16 fb$^{-1}$}
    \begin{itemize}
    \item Three central ($|\eta|<1$) photons, $E_T > 13$ GeV
    \item Isolation: $E_T < 2$ GeV inside a cone of 0.4 in $\eta-\phi$ space
    \end{itemize}
\section{Selection for LHC and future colliders\label{ss:future_tri}}
\subsection{Future $e^+e^-$ colliders}
    Tri-$\gamma$ selection for future $e^+e^-$ colliders as a function of $M_a$.
    \begin{itemize}
        \item Angular acceptance, three photons within $40^\circ < \theta < 140^\circ$
        \item Isolation criteria: angular separation no less than 15$^\circ$
        \item One photon with energy $\Big|E_\gamma-  \frac{s-M_a^2}{2\sqrt{s}}\Big| < E_c$ GeV
        \item Remaining two photons with invariant mass $|M_{\gamma\gamma}-M_a| <  E_c$ and angular separation $\theta_\gamma < \delta\theta < \theta_\gamma+20^\circ$
        \item $\cos\theta_\gamma \equiv -\frac{4s M_a^2}{(s+M_a^2)^2}$
    \end{itemize}
    \begin{center}
    \begin{tabular}{cccc}
        Collider&$\sqrt{s} $ [GeV]&$\mathcal{L}_{\text{int.}}$[ab$^{-1}$]& $E_{c}$ [GeV]\tabularnewline\hline
        Belle II& 10.6& 50& 0.1\tabularnewline
        ILC& 240& 1& 5\tabularnewline
        TLEP& 1000& 10& 10
        \end{tabular}
        \end{center}

\subsection{LHC at 8 and 13 TeV}
    Tri-$\gamma$ selection for LHC analyses as a function of $M_a$. Signal and irreducible tri-photon signals were generated at leading order using {\sc CTEQ6L} PDF set while the reducible diphoton + jet background was generated by simulating the diphoton process at NLO using the {\sc NNPDF23\_nlo\_as\_0118\_qed} PDF sets and was showered using {\sc PYTHIA} 8~\cite{Sjostrand:2007gs}. The analyses for 8 and 13 TeV were identical up to the basic acceptance requirements and mass dependent energy cut on the recoiling photon.
 \subsubsection*{Vetos}
    \begin{itemize}
        \item More than one jet with $p_T > p_T^{\text{acc.}}$ 
    \end{itemize}
 \subsubsection*{Selection}
\begin{itemize}
    \item Three isolated photons (isolation cone of radius $\Delta R$=0.3) within $|\eta| < 2.5$ and with $p_T > p_T^{\text{acc.}}$
    \item Two photons with invariant mass closest to $M_a$ required to have $|M_{\gamma\gamma}-M_a| < 0.1M_a$
    \item Energy of the remaining `recoil' photon in the three photon centre of mass frame should have $E_{\text{rec.}} > \bar{E}_{\text{rec.}}$
\end{itemize}
\begin{center}
    \begin{tabular}{ccc}
    $\sqrt{s}$ (TeV)& $p_T^{\text{acc.}}$ (GeV)&$\bar{E}_{\text{rec.}}$ (GeV)\tabularnewline\hline
    8&20& $M_a/6 + 70. $\tabularnewline
    13&30& $\text{max}(100.,\,M_a/5 + 80.) $\tabularnewline
    \end{tabular}
\end{center}

\section{CMS and ATLAS monojet selections\label{AppendixC}}
A summary of the main criteria of two LHC monojet analyses used to set limits on the ALP mass and its coupling to gluons. See Refs.~\cite{CMS:rwa,ATLAS:2012zim} for more details.
\subsection{CMS}
\subsubsection*{Trigger}
\begin{center}
    \begin{minipage}{0.45\linewidth}
        \centering
    Option 1
    \begin{itemize}
    \item $\slashed{E}_T>120$ GeV 
    \end{itemize}
    \phantom{AAA}
\end{minipage}
\begin{minipage}{0.45\linewidth}
\begin{center}
Option 2
    \begin{itemize}
    \item 1 jet with $p_T>80$ GeV $|\eta| < 2.6$
    \item $\slashed{E}_T > 105$ GeV
    \end{itemize}
\end{center}
\end{minipage}
\end{center}

\subsubsection*{Vetos}
    \begin{itemize}
    \item More than two jets with $p_T > 30$ GeV and $|\eta| < 4.3$ 
    \item A second jet with $\Delta\phi(j_1,j_2)> 2.5$ 
    \item Well reconstructed electrons or muons with $p_T > 10$ GeV 
    \item Well reconstructed taus with $p_T > 20$ GeV and $|\eta| < 2.3$
    \end{itemize}
\subsubsection*{Selection}
    \begin{itemize}
    \item Hardest jet with $p_T > 110$ GeV and $|\eta| < 2.4$ 
    \item $\slashed{E}_T$ cut for each signal region
    \end{itemize}

\begin{center}
 \begin{tabular}{|c|c|c|c|c|c|c|c|}
     \hline
     Signal region            & SR1 & SR2 & SR3 & SR4 & SR5 & SR6 & SR7 \\ \hline
     $\slashed{E}_T$ cut [GeV]& 250 & 300 & 350 & 400 & 450 & 500 & 550 \\ \hline
 \end{tabular}       
\end{center}

\subsection{ATLAS}
\subsubsection*{Trigger}
    \begin{itemize}
    \item $\slashed{E}_T>80$ GeV 
    \end{itemize}
\subsubsection*{Vetos}
    \begin{itemize}
    \item Any jet with $p_T > 20$ GeV and $|\eta| < 4.5$  with anomalous charge fraction, electromagnetic fraction in the calorimeter or timing
    \item More than two jets with $p_T > 30$ GeV and $|\eta| < 4.5$ 
    \item A second jet with $\Delta\phi(\slashed{E}_T,j)> 0.5$ 
    \item Reconstructed electrons with $p_T > 20$ GeV $|\eta| < 2.47$
    \item Reconstructed muons with $p_T > 7$ GeV $|\eta| < 2.5$
    \end{itemize}
\subsubsection*{Selection}
    \begin{itemize}
    \item $\slashed{E}_T > 120$ GeV 
    \item at least one jet with $p_T > 120$ GeV and $|\eta| < 2$ 
    \item Additional $\slashed{E}_T$ and leading jet $p_T$ cut for each signal region
    \end{itemize}
    
\begin{center}
 \begin{tabular}{|c|c|c|c|c|}
     \hline
     Signal region              & SR1 & SR2 & SR3 & SR4  \\ \hline
     leading jet $p_T$ cut [GeV]& 120 & 220 & 350 & 500 \\ \hline
     $\slashed{E}_T$ cut [GeV]  & 120 & 220 & 350 & 500 \\ \hline
 \end{tabular}       
\end{center}

 \end{document}